\documentclass[twocolumn, amsmath, amssymb, floatfix]{revtex4}

 \usepackage{amsmath,mathrsfs,graphicx,epstopdf,placeins,float}
 \usepackage{booktabs}
 \usepackage{multirow}

\begin{document}

 \title{ Dynamics of  domain wall in charged AdS  dilaton black hole spacetime  }
 \author{Wu-Long Xu}
 \email{wlxu@emails.bjut.edu.cn}
 
 \author{Yong-Chang Huang}
 \email{ychuang@bjut.edu.cn}
 \affiliation{Institute of Theoretical Physics, Beijing University of Technology, Beijing 100124, China}
 \begin{abstract}
 For the $n-1$ dimensional  FRW domain wall universe induced by  $n$ dimensional charged dilaton black hole, its movement formula  in the bulk can be rewrite as the expansion or collapsing of domain wall. By analysing, we found that in this static AdS space, the cosmologic behaviour of  domain wall is particularly single. Even more surprising, it exists an anomaly that the domain wall has a motion area outside of horizon, in which it cannot be explained by our classical theory.
 \end{abstract}
 \maketitle
\section{Introduction}
As we know that general relativity (GR) is not only a beautiful theory but also consistent with the observations.  But there are  still many problems that GR can't solve \cite{Will:2014kxa}.  Now we can not define which inflation model is the most accurate \cite{Martin:2013tda} and it also includes the problem of dark matter and dark energy \cite{Jungman:1995df}-\cite{Ryskin:2014pva}.  Thus other gravity theory is coming into being such as $f(R)$ gravity \cite{DeFelice:2010aj}, Brans-Dicke gravity \cite{Gao:2006dx},  higher-order gravitational theory \cite{Maeda:2003vq},  dilaton gravity \cite{Garfinkle:1990qj}  and so on \cite{Prabhu:2018aun}, \cite{Sotiriou:2015lxa}.

Among these theories beyond GR, dilaton gravity is regarded as the low energy limit of string theory.  Adding dilaton field will take many interesting characters such as dilaton charged black hole. Ref.  \cite{Chan:1995fr}  got a calss metric solutions that is neither asympototically flat nor (A)dS . These solutions just have a singularity on the origin.  Also when the dilaton potential is token as Liouville type,  People can gain different  black hole solutions \cite{Agnese:1994zx}-\cite{Sheykhi:2014ipa}.  Then C.~J.~Gao and S.~N.~Zhang present the AdS  metric solution in the charged dilaton black hole  \cite{Gao:2004tv}, \cite{Gao:2005xv}. 

 The AdS/CFT duality originated from superstring theory is one of  the best applications for the AdS spacetime \cite{Natsuume:2014sfa}.  And the  holography in the charged dilaton black hole have been studied deeply \cite{Zhou:2018ony}-\cite{Zhang:2015dia}.

Because of the speciality and important of AdS space \cite{Gibbons:2011sg}, in which many physics have been discussed \cite{Li:2017kkj}, it is necessary to study the motion of  domain wall in the charged AdS dilaton black hole bulk now \cite{Christensen:1998hg}, \cite{Frolov:1998td}. Indeed, we find that its motion is single and  it exists an anomaly that is not explained by our classical theory. 

Our paper is arranged as follow. In section II, we introduce the charged dilaton black hole bulk. Its metric solution shows the (A)dS characater. In section III,  we describle a bulk-wall system in which, the Friedmann equations on the domain wall  can be induced by the Israel matching conditions. Thus we can rewrite the equations of motion of the domain wall in the bulk  as the expression of expansion or collapsing of that. In the section IV, we specifically analyze the motion of domain wall by the speed-coordinates diagram. It found that in this AdS space, the motion of domain wall is single. Meanwhile it exists an anomaly. In section V, we make the conclusions and discussion.
\section{charged  dilaton black hole bulk}
In the theory  of  $n$ dimensional   Einstein-Maxwell-dilaton gravity, the action is 
\begin{equation}\label{action}
S_{bulk}=\int dx^{n}\sqrt{g}[R-\frac{4}{n-2}(\nabla\phi)^{2}-V(\phi)-e^{\frac{-4}{n-2}\alpha\phi}F^{MN}F_{MN}],
\end{equation}
where $R$ is the Ricci scalar, $\phi$ and V($\phi$) is the dilaton field and dilaton potential.
$F_{MN}=\partial_{M}A_{N}-\partial_{N}A_{M}$ is the electromagnetic tentor. The constant $\alpha$ denotes the coupling strength between the Maxwell field and the dilaton field.  The potential function is taken as
\begin{equation}\label{potentialfunction}
\begin{split}
V(\phi)=&\frac{2\Lambda}{(n-1)(n-3+\alpha^{2})}\\
&(-\alpha^{2}(n^{2}-n\alpha^{2}-6n+\alpha^{2}+9)e^{-\frac{4(n-3)\phi}{(n-2)\alpha}}\\
&+(n-3)^{2}(n-1-\alpha^{2})e^{\frac{4\alpha\phi}{n-2}}\\
&+4\alpha^{2}(n-2)(n-3)e^{-2\frac{n-3-\alpha^{2}}{(n-2)\alpha}}),
\end{split}
\end{equation}
where $\Lambda$ is cosmology constant and $\Lambda=-\frac{(n-1)(n-2)}{2l^{2}}$.

Varing action (\ref{action}) with respect to the metric $g^{AB}$, dilaton field $ \phi$, Maxwell field $A^{M}$, respectively,  the dynamic equations can be gotten.
\begin{eqnarray}
&&R_{MN}=\frac{4}{n-2}(\partial_{M}\phi\partial_{N}\phi+\frac{1}{4}g_{MN}V)\nonumber\\
&&~~~~~~~~~+2e^{-\frac{4}{n-2}\alpha\phi}(F_{MP}F_{N}^{P}-\frac{1}{2(n-2)}g_{MN}F^{2})\label{Einsteinfieldequations}\\
&&\nabla^{2}\phi=\frac{n-2}{8}\frac{\partial V}{\partial\phi}-\frac{\alpha}{2}e^{-\frac{4\alpha\phi}{n-2}F^{2}}\\
&&\nabla_{M}(\sqrt{-g}e^{-\frac{4\alpha\pi}{n-2}}F^{MN})=0.	
\end{eqnarray}

The form of metric is 
\begin{equation}\label{metric}
ds^{2}=-N^{2}(r)f^{2}(r)dt^{2}+\frac{dr^{2}}{f^{2}(r)}+R^{2}(r)d\Omega_{k,n-2}^{2},
\end{equation}
where  $d\Omega_{n-2}^{2}$ is the line element of $n-2$ dimensional space with constant scalar curvature$(n-2)(n-3)k$.
$k=0,\pm1$ donates the topology of horizon as to be plane, sphheric and hyperbolic respectively. 

Taking this metric (\ref{metric}) into the Maxwell field equation, it will have 
\begin{equation}
F_{tr}=N(r)\frac{qe^{\frac{4\alpha\phi}{n-2}}}{R^{n-2}},
\end{equation}
where we just conside the Maxwell field as electric charge and $q$ is the charge of black hole.

Using the metric and the dilaton potential function, these field equations can be solved as (the detail can be found by the Ref. \cite{Gao:2005xv}.
\begin{eqnarray}\label{solutions}
&&N^{2}(r)=(1-(\frac{b}{r})^{n-3})^{-\gamma(n-4)}\\
&&f^{2}(r)=(1-(\frac{b}{r})^{n-3})^{\gamma(n-4)}[(k-(\frac{c}{r})^{n-3})\nonumber\\
&&~~~~~~~~~~\times(1-(\frac{b}{r})^{n-3})^{1-\gamma(n-3)}
-\frac{2\Lambda}{(n-1)(n-2)}r^{2}\nonumber\\
&&~~~~~~~~~~\times(1-(\frac{b}{r})^{n-3})^{\gamma}]\\
&&R^{2}(r)=r^{2}(1-(\frac{b}{r})^{n-3})^{\gamma}\\
&&\gamma=\frac{2\alpha^{2}}{(n-3)(n-3+\alpha^{2})}\\
&&\phi(r)=\frac{(n-2)\alpha}{2(n-3+\alpha^{2})}\ln(1-(\frac{b}{r})^{n-3}),
\end{eqnarray}
where $b$ and $c$ are integration constants with the length of the dimension.  The charge $q$ related with $b$ and $c$ is 
\begin{equation}\label{charge}
q^{2}=\frac{(n-2)(n-3)^{2}}{2(n-3+\alpha^{2})}b^{n-3}c^{n-3}.
\end{equation}

\section{the dynamic equation of  domain wall}
For a domain wall-bulk system, $n-1$ dimensional domain wall is embedded into the bulk in parallel. Here we want to take $n=5$ for explaining the physics.  Firstly,   we  take the position  of one  object  $X^{A}(t,r,x,y,z)$ in the bulk coordinate, while it is $x^{\mu}(\tau,x,y,z)$ in the domain wall coordinate \cite{Xu:2019gzt}. The domain wall move along with the coordinate $r$ in the bulk.  When we consider the domain wall as our universe, it shows four dimensional FRW spacetime can be induced by the five dimensional bulk metric, namely, $X^{A}(x^{\mu})=(t(\tau),r(\tau),x,y,z)$.  Next we will get it.  We ansatz the  generally static metric as 
\begin{equation}\label{staticmetric}
ds^{2}=g_{AB}dx^{A}dx^{B}=-A(r)dt^{2}+B(r)dr^{2}+R(r)^{2}d\Omega^{2}.
\end{equation} 
The velocity of object is $V^A(\dot{t},\dot{r},0,0,0)$. The ovedots and primes is differentiation with respect to $\tau$ and $r$  respectively in the next place. According to the condition $g_{AB}V^{A}V^{B}=-1$, it can be obtained that is 
\begin{equation}\label{deformation}
-A\dot{t}^{2}+B\dot{r}^{2}=-1\rightarrow\dot{t}=\sqrt{\frac{B\dot{r}^{2}+1}{A}}.
\end{equation}
Thus, the form of induced FRW metric is 
\begin{equation}\label{inducedmetric}
ds^{2}=-d\tau^2+R(r(\tau))^{2}d\Omega^{2}.
\end{equation}
Then it should be know that the  extrinsic curvature for the wall is 
\begin{equation}\label{curvature}
K_{AB}=h_{A}^{M}h_{B}^{N}\nabla_{M}n_{N},
\end{equation}
where $n^A(m,l,0,0,0)$ is the normal vectora and $h_{AB}$ is the induced metric of domain wall.  

Using the conditions 
\begin{eqnarray}\label{conditions}
&& V_{A}n^{A}=0\\
&&n^{B}n^{A}g_{AB}=1,
\end{eqnarray}
we have the expression of $n_A$
\begin{equation}\label{na}
n_{A}=(\sqrt{AB}\dot{r},-\sqrt{B(1+B\dot{r}^{2})},0,0,0)
\end{equation}
Thus the components of  extrinsic curvature can be written as 
\begin{eqnarray}
&&K_{ij}=\nabla_{i}n_{j}=\frac{h_{ij}}{R}R'\sqrt{\frac{1+B\dot{r}^{2}}{B}}\label{components1}\\
&&K_{\tau\tau}=\frac{dt}{d\tau}\frac{dt}{d\tau}K_{tt}=\frac{1}{AB}\frac{d}{dr}(\sqrt{A(1+B\dot{r}^{2})})\label{components2}.
\end{eqnarray}

In the $n-1$ dimensional domain wall, the action is 
\begin{equation}\label{dwaction}
S_{dw}=\int dx^{4}\sqrt{-h}\{K\}+\int dx^{4}\sqrt{-h}\mathscr{L}_{M}(\tilde{h}_{\mu\nu}),
\end{equation}
where $K=h^{AB}K_{AB}$, $\mathscr{L}_{M}$ the matter on the domain wall and we do the conform transform for the $h_{\mu\nu}$ \cite{Brax:2002nx},
\begin{equation}\label{transform}
\tilde{h}_{\mu\nu}=e^{2\eta{\phi}}h_{\mu\nu}.
\end{equation}
The total action of system  is 
\begin{equation}\label{totalaction}
S=S_{bulk}+S_{dw}.
\end{equation}
Varying the action (\ref{totalaction}) with respect to the metric $g^{AB}$ and the scalar field $\phi$, the field equations on the domain wall can be written as 
\begin{eqnarray}
&&\{K_{AB}-Kh_{AB}\}=-(-2\frac{\delta\mathscr{L}_{M}}{\delta h^{AB}}+h_{AB}\mathscr{L}_{M})\label{dwfieldequations1}\\
&&\frac{4}{3}{n\cdot\partial\phi}=\frac{d\mathscr{L}_{M}}{d\phi}\label{boundryequationphi}.
\end{eqnarray}
Here we will assume that the domain wall has a $Z_2$ symmetry and the $\{\}_{-}$ wil be used to calculate. When we consider the matter as the perfect fluid on the domain wall, the energy tensor $T^{\mu}_{\nu}=Dig(-\rho,P,P,P)$.
Expanding Eq. (\ref{Einsteinfieldequations}), when $n=5$, we can acquire a equation.
\begin{equation}
\frac{1}{\sqrt{AB}}\frac{d}{dr}(\sqrt{AB}\frac{R}{R'})=1+\frac{4R^{2}(\phi')^{2}}{9(R')^{2}}.\label{additioncondition}
\end{equation}
Taking the Eqs. (\ref{components1}, \ref{components2}, \ref{additioncondition}) into the Eqs. (\ref{dwfieldequations1}, \ref{boundryequationphi}), the boundry equations on the domain wall are
\begin{eqnarray}
&&\sqrt{\frac{1+B\dot{r}^{2}}{B}}=\sqrt{\frac{1}{B}+\dot{r}^{2}}=\frac{R}{R'}\frac{1}{6}\rho\label{boundryequation1}\\
&&\phi'=\frac{9}{4}\frac{R'}{R}\xi'(3w-1)\label{boundryequation2}\\
&&\dot{\rho}+3H(P+\rho)=-\frac{4}{9}\frac{\rho\dot{\phi}^{2}}{H}\label{boundryequation3},
\end{eqnarray}
where the Habble constant $H=\frac{\dot{R}}{R}=\frac{R'\dot{r}}{R}$ and $w=\frac{P}{\rho}$.
Taking the Eq. (\ref{boundryequation2}) and the bulk metric (\ref{metric})  into Eq. (\ref{boundryequation3}), the expression of density $\rho$ is 
\begin{equation}\label{density}
\begin{split}
\rho&=\rho_{0}+r^{-\frac{2(3+3w+\alpha^{2})}{2+\alpha^{2}}}(r^{2}-b^{2})^{-\frac{3}{2}(1+w)+\frac{1+3w}{2+\alpha^{2}}}\\
&\times(-2b^{2}+r^{2}(2+\alpha^{2})).
\end{split},
\end{equation}
where $\rho_0$ is a constant.
Combining Eqs. (\ref{metric}, \ref{density}) with Eq. (\ref{boundryequation1}),  the dynamic equation of domain wall is 
\begin{equation}\label{dynamicequation}
\dot{R}^{2}=V(r)=(\frac{1}{6}\rho R )^{2}-(fR')^{2}.
\end{equation}
We will take metric into Eq. (\ref{dynamicequation}), it has 
\begin{equation}\label{potentialfunction0}
\begin{split}
V(r)=&\frac{1}{(b^2-r^2)^2(2+\alpha^2)^2}(1-\frac{b^2}{r^2})^{\frac{\alpha^2}{2+\alpha^2}}(-2b^2+r^2(2+\alpha^2))^2\\
&\times(r^{\frac{-4}{2+\alpha^2}}(r^2-b^2)^{\frac{2}{2+\alpha^2}}(\frac{c^2}{r^2}-k)\\
&+\frac{1}{6}r^{\frac{4-2\alpha^2}{2+\alpha^2}}(r^2-b^2)^{\frac{2\alpha^2}{2+\alpha^2}}\Lambda\\
&+\frac{1}{36}r^{-\frac{2(4+6w+\alpha^{2})}{2+\alpha^{2}}}(r^{2}-b^{2})^{-1+\frac{2+6w}{2+\alpha^{2}}-3w}(2+\alpha^{2})^{2}).
\end{split}
\end{equation} 
Here it is the process of potential energy into kinetic energy. Finally Potential energy $V(r)$ will become to be kinetic energy in the infinity.
\section{The evolution process}
Here we get the evolution equation (\ref{dynamicequation}) of domain wall, it indicates that $V(r)$ must be big than zero. Because of the complexity of $V(r)$, it is difficult that knows its montonicity by analyzing these parameters $w, \alpha$ in the all range of $r$. But using the increasing character of exponential functions, we can know the laws of  $V(r)$ in the $r>1$ easily in the  range of different parameters. Thus when we scan the all parameters range in the $r>1$ in the picture, once the  parameters are definite, the all image of $V(r)$ is clear in the $0<r<\infty$.  Then according the speed-position pictures, the evolution of domain wall along  with the time can be shown in the different topology.
Meanwhile we want to know the more characates of  AdS space, so that $\Lambda=-1$. In fact we know from the follow contents that  in the dS and flat space, the motion of doain wall is expansion in the whole motor  process.

Type I: $k=1$. 

Case (i): If $r>\frac{1}{2}(1+\sqrt{1+4b^2})>c$,  then $r^2-b^2>r>1$.  Thus the expression of $V(r)$ can be written as approximatively
\begin{equation}\label{potentialfunctionk1}
\begin{split}
V(r)=&\frac{1}{(b^2-r^2)^2(2+\alpha^2)^2}(1-\frac{b^2}{r^2})^{\frac{\alpha^2}{2+\alpha^2}}(-2b^2+r^2(2+\alpha^2))^2\\
&\times(-r^{\frac{-4}{2+\alpha^2}}(r^2-b^2)^{\frac{2}{2+\alpha^2}}\\
&-\frac{1}{6}r^{\frac{4-2\alpha^2}{2+\alpha^2}}(r^2-b^2)^{\frac{2\alpha^2}{2+\alpha^2}}\\
&+\frac{1}{36}r^{-\frac{2(4+6w+\alpha^{2})}{2+\alpha^{2}}}(r^{2}-b^{2})^{-1+\frac{2+6w}{2+\alpha^{2}}-3w}(2+\alpha^{2})^{2}).
\end{split}
\end{equation}
Due to the montonicity of function, $r^{\frac{2\alpha^2-8}{2+\alpha^2}}<(r^2-b^2)^{\frac{2\alpha^2-2}{2+\alpha^2}}$. 
Compairing the first and the second terms, its deformation is 
\begin{equation}\label{deformation1}
\begin{split}
&-r^{\frac{-4}{2+\alpha^2}}(r^2-b^2)^{\frac{2}{2+\alpha^2}}+r^{\frac{4-2\alpha^2}{2+\alpha^2}}(r^2-b^2)^{\frac{2\alpha^2}{2+\alpha^2}}=\\
&r^{\frac{-4}{2+\alpha^2}}(r^2-b^2)^{\frac{2}{2+\alpha^2}}(r^{\frac{8-2\alpha^2}{2+2\alpha^2}}(r^2-b^2)^{\frac{2\alpha^2-2}{2+\alpha^2}}-1)>0.
\end{split}
\end{equation}
Similarly, if $V(r)$  is larger than zero, it needs
\begin{equation}\label{condition1}
(r^2-b^2)^{-1+\frac{2+6w-2\alpha^2}{2+\alpha^2}-3w}>r^{\frac{2(4+6w+\alpha^2)+4-2\alpha^2}{2+\alpha^2}}. 
\end{equation}
 Thus we have 
\begin{equation}\label{deformation2}
\begin{split}
&r^{\frac{2\alpha^2}{2+\alpha^2}}(r^2-b^2)^{\frac{2\alpha^2}{2+\alpha^2}}(r^{-\frac{2(4+6w+\alpha^2)+4-2\alpha^2}{2+\alpha^2}}\\
&\times(r^2-b^2)^{-1+\frac{2+6w-2\alpha^2}{2+\alpha^2}-3w}-1)>0.
\end{split}
\end{equation}
Then using the character of exponential function and the Eq. (\ref{condition1}) the range of parameters $w, \alpha$ can be determinated as $w<1, \alpha\in R$.  For the range $0<r<\frac{1}{2}(1+\sqrt{1+4b^2})$, In fact, we don't needs analyse its parameter. Because once we scan the parameters in the $r>\frac{1}{2}(1+\sqrt{1+4b^2})>c$, we can read its trend form the pictures. 
\begin{figure}[!h]
\includegraphics[width=4.2cm]{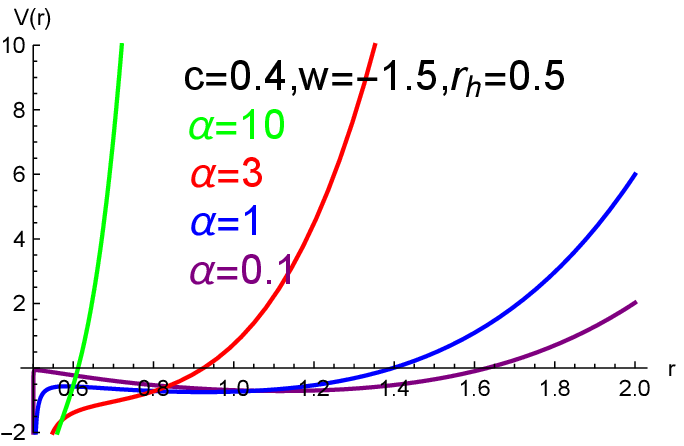}
\includegraphics[width=4.2cm]{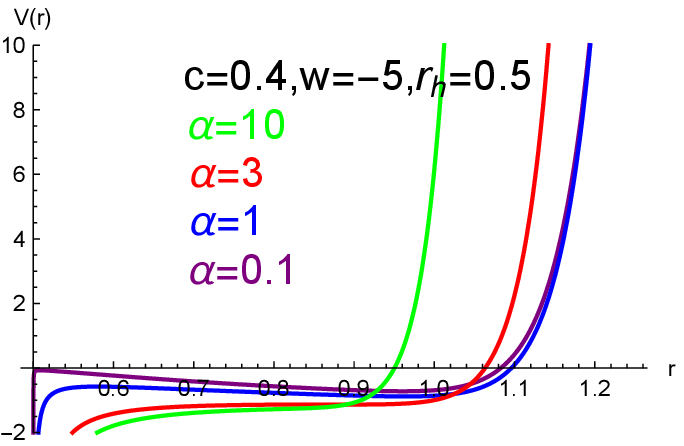}
\caption{$V(r)$ versus $r$ for $ k=1$.}
\label{picture1}
\end{figure}

Suprisedly, when $k=1$, domain wall will exist an abnormal  region that the potential is negative out of the horizon. Thus we can not explain this anomaly using classcial physics. We guess that the domain wall should exist a quantum behaviour, such as tunnel effect. The transmlss probability is decided by the energy of domain wall and the width of tunnel.  Here we mainly want to study the motion of domain wall. So this anomaly will be researched in our next paper. 

In a ward, we  will find that if $\dot{R}>0$,  the domain wall firstly goes through an anomaly then accelerative expansion   from the FIG. (\ref{picture1}). If $\dot{R}<0$, the domain wall collpases until it stops  outside of the horizon.

Case (ii): If $r>c>\frac{1}{2}(1+\sqrt{1+4b^2})>b$, we still plan to take the way that for the big range $c<r<\infty$ we use  the monotonous character of exponential function to scan the all range of parameters satisfied with $V(r)>0$; for the rest range of $r$, we can look the functional pictures  directly. So its result is same to the above scenario. Its functional images are following that
\begin{figure}[!h]
	\includegraphics[width=4.2cm]{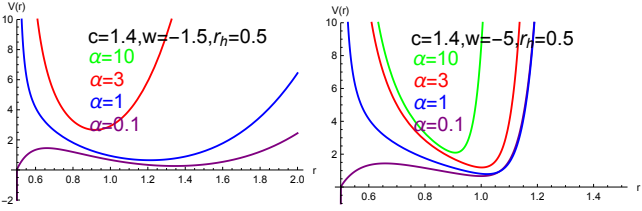}
	\includegraphics[width=4.2cm]{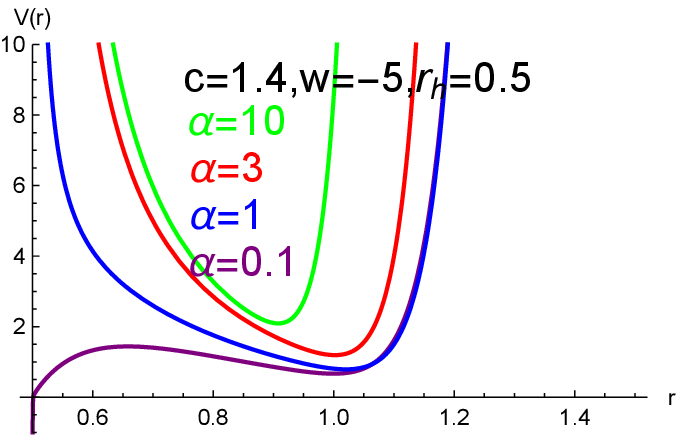}
	\caption{$V(r)$ versus $r$ for $ k=1$.}
	\label{picture2}
\end{figure}
 
In this case, it indicates that the domain wall is firstly  slow expansion reached a minimal  value then accelerative  expansion when the coupling constant is not small. From the FIG. (\ref{picture2}), we can see that when the dilaton conpling constant is very samll, the motion of  domain wall is very different. It is firstly accelerative expansion reached a maximum value then decreasing expansion until a minimal value followed by accelerative expansion again.

Type II: $k=0$. 

Case (i): The expression of $V(r)$ is 
\begin{equation}\label{potentialfunctionk2}
\begin{split}
V(r)=&\frac{1}{(b^2-r^2)^2(2+\alpha^2)^2}(1-\frac{b^2}{r^2})^{\frac{\alpha^2}{2+\alpha^2}}(-2b^2+r^2(2+\alpha^2))^2\\
&\times(r^{\frac{-8-2\alpha^2}{2+\alpha^2}}(r^2-b^2)^{\frac{2}{2+\alpha^2}}c^2\\
&-\frac{1}{6}r^{\frac{4-2\alpha^2}{2+\alpha^2}}(r^2-b^2)^{\frac{2\alpha^2}{2+\alpha^2}}\\
&+\frac{1}{36}r^{-\frac{2(4+6w+\alpha^{2})}{2+\alpha^{2}}}(r^{2}-b^{2})^{-1+\frac{2+6w}{2+\alpha^{2}}-3w}(2+\alpha^{2})^{2}).
\end{split}
\end{equation}
We take the range $r>\frac{1}{2}(1+\sqrt{1+4b^2})$ to scan the parameters space. Because of $(r^2-b^2)^{2-2\alpha^2}<r^12$,  it has
\begin{equation}\label{deformationk0}
\begin{split}
&r^{\frac{-8-2\alpha^2}{2+\alpha^2}}(r^2-b^2)^{\frac{2}{2+\alpha^2}}c^2
-\frac{1}{6}r^{\frac{4-2\alpha^2}{2+\alpha^2}}(r^2-b^2)^{\frac{2\alpha^2}{2+\alpha^2}}=\\
&r^{\frac{4-2\alpha^2}{2+\alpha^2}}(r^2-b^2)^{\frac{2\alpha^2}{2+\alpha^2}}(r^{\frac{-12}{2+\alpha^2}}(r^2-b^2)^{\frac{2-2\alpha^2}{2+\alpha^2}}-1)<0.
\end{split}
\end{equation}
Thus if $V(r)>0$, it needs to compare the second and the third term. its result can be acqiured by the first scenairo. So the range of parameters are $w<1, \alpha\in R$. In this situation, the image is as follow.
\begin{figure}[!h]
	\includegraphics[width=4.2cm]{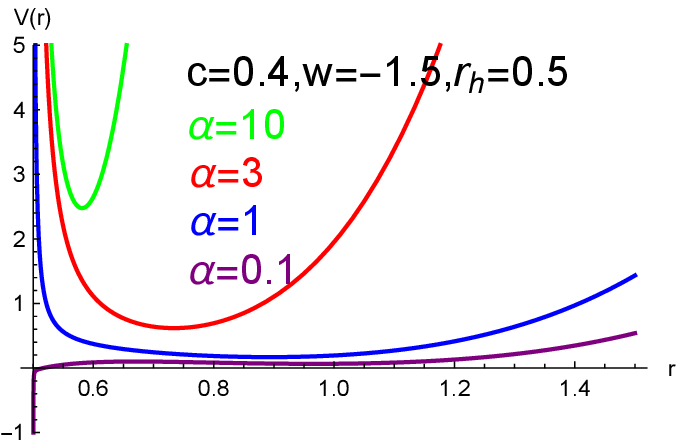}
	\includegraphics[width=4.2cm]{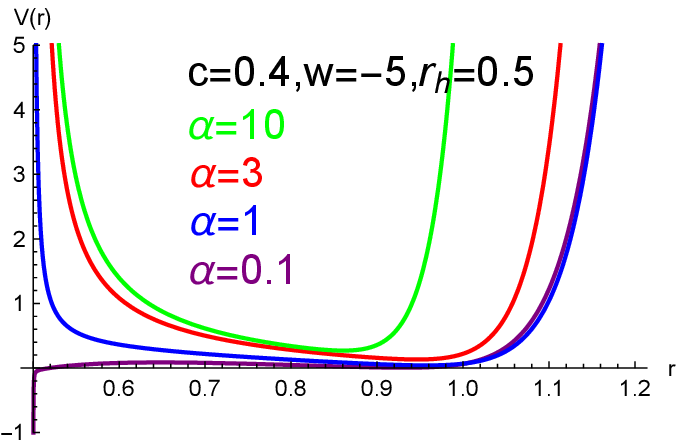}
	\caption{$V(r)$ versus $r$ for $ k=0$.}
	\label{picture3}
\end{figure}	

Obvering the FIG. (\ref{picture3}), it can be found that this situation is similar to the case (ii) of $k=-1$.  But it is diferent that when $c$ is small, the length of  jump from decrreasing accelerative expansion  to  accelerative expansion is more longer.

Type III: $k=-1$. 

\begin{figure}[!h]
	\includegraphics[width=4.2cm]{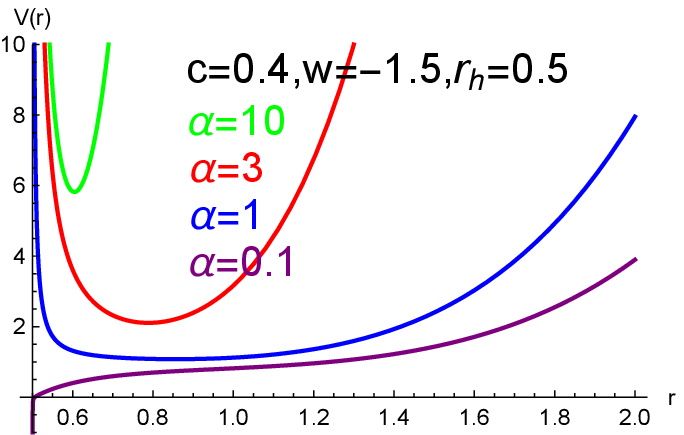}
	\includegraphics[width=4.2cm]{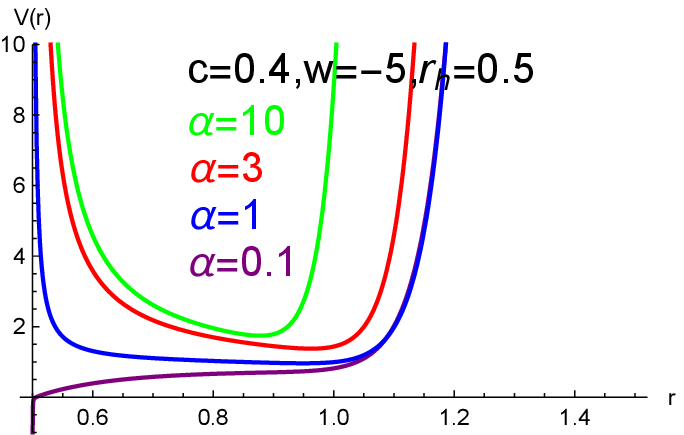}
	\caption{$V(r)$ versus $r$ for $ k=-1$.}
	\label{picture4}
\end{figure}

Case (i): If $r>\frac{1}{2}(1+\sqrt{1+4b^2})>c$.  The form of $V(r)$ is 
\begin{equation}\label{potentialfunctionk3}
\begin{split}
V(r)=&\frac{1}{(b^2-r^2)^2(2+\alpha^2)^2}(1-\frac{b^2}{r^2})^{\frac{\alpha^2}{2+\alpha^2}}(-2b^2+r^2(2+\alpha^2))^2\\
&\times(r^{\frac{-4}{2+\alpha^2}}(r^2-b^2)^{\frac{2}{2+\alpha^2}}\\
&-\frac{1}{6}r^{\frac{4-2\alpha^2}{2+\alpha^2}}(r^2-b^2)^{\frac{2\alpha^2}{2+\alpha^2}}\\
&+\frac{1}{36}r^{-\frac{2(4+6w+\alpha^{2})}{2+\alpha^{2}}}(r^{2}-b^{2})^{-1+\frac{2+6w}{2+\alpha^{2}}-3w}(2+\alpha^{2})^{2}).
\end{split}
\end{equation}
It has $(r^2-b^2)^{2-2\alpha^2}<r^{-2\alpha^2+8}$, Thus 
\begin{equation}\label{deformation3}
r^{\frac{4-2\alpha^2}{2+\alpha^2}}(r^2-b^2)^{\frac{2\alpha^2}{2+\alpha^2}}(r^{\frac{2\alpha^2-8}{2+\alpha^2}}(r^2-b^2)^{\frac{2-2\alpha^2}{2+\alpha^2}}-1)<0.
\end{equation}
If $V(r)>0$, we have
\begin{equation}\label{condition3}
r^{-\frac{-2(4+6w+\alpha^2)}{2+\alpha^2}}(r^2-b^2)^{-1+\frac{2+6w}{2+\alpha^2}-3w}-r^{\frac{4-2\alpha^2}{2+\alpha^2}}(r^2-b^2)^{\frac{2\alpha^2}{2+\alpha^2}}>0
\end{equation}
So that we can derive the range of  parameters $w<1, \alpha\in R$. Then the functional picture is FIG. (\ref{picture4}) in the all value of $r$.

\begin{figure}[!h]
	\includegraphics[width=4.2cm]{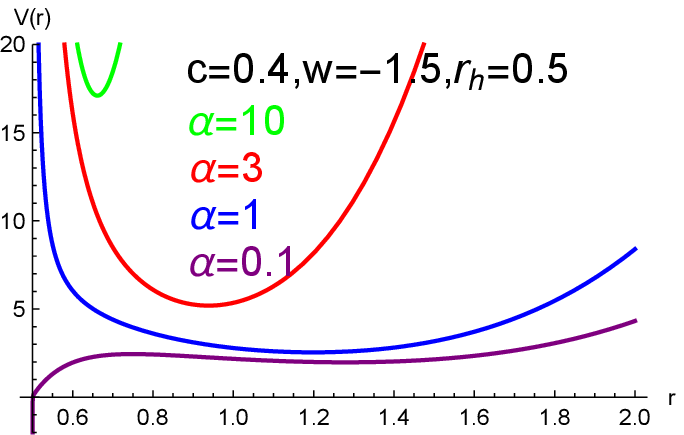}
	\includegraphics[width=4.2cm]{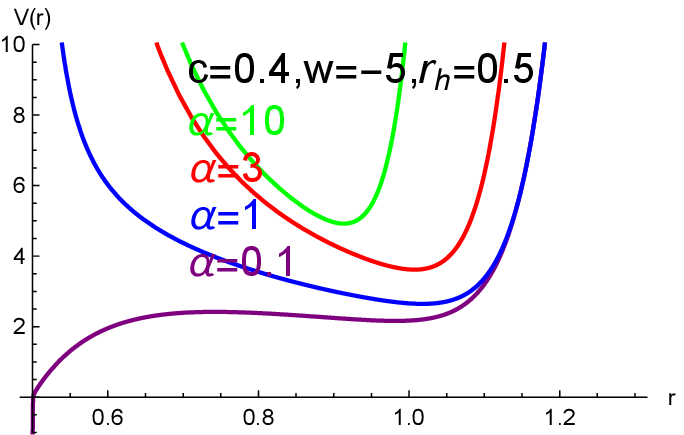}
	\caption{$V(r)$ versus $r$ for $ k=-1$.}
	\label{picture5}
\end{figure}

Case (ii): If $r>c>\frac{1}{2}(1+\sqrt{1+4b^2})>b$, we adopt the first scenairo.  Its functional picture is FIG. (\ref{picture5}).

For $k=-1$, it shows that for this two cases the motion of domain wall  is same to the previous examples such as $k=0$ from their pictures.  

Generally, in a AdS space, if the backgroud is the static charged dilato black hole,  the motion of  domain wall  is always firstly  decreasing expansion reached a minimal value then accelerative expansion  in the bigger value's coupling constant. When the coupling constant is smaller, the motion of domain wall is firstly accelerative expansion reached a maximum value then decreasing expansion until a minimal value followed by accelerative expansion again.  Thus its motion is pretty single. But it exists an anomaly when $k=-1$ and $c$ is smaller. For this case, we can not use the classical thery to explain it. 
\section{conslusions and discussion}
In the AdS space,  the $n-1$ dimensional  FRW domain wall universe  can be induced by  the $n$ dimensional dilaton charged  black hole bulk.  We use the  metric  solved by the Gao and Zhang  \cite{Gao:2005xv} to analyse the cosmologic behaviour of domain wall. 
By the Israel matching conditions, we can rewrite the motion of domain wall in the bulk as the expansion and collapse of domain wall. In the course of analysis, we use a novel way that on the larged region of motion, we use the montonicity of exponential function to define the parameters range satisfying the condition $V(r)>0$. Then by scaning all parameters range, the motion images can be made easily in the whole area. We choose three different topologies of horizon to observe the behaviour of domain wall. It is found that the motion of domain wall is single that when the coupling constant is bigger, the motion of domain wall  is  firstly  decreasing expansion reached a minimal value then accelerative expansion; when the constant is smaller, its motion  is   firstly  accelerative expansion reached a maximum value then decreasing expansion until a minimal value followed by accelerative expansion again.  Also the length of  first stage of accelerative expansion and then descrising expansion is determined by the parameters $w,\alpha, c$. But it is excited that when $k=1$ and $\frac{1}{2}(1+\sqrt{1+4b^2})>c$, we found it exists an anomaly that the potential energy of domain wall is negative in certain location outside of horizon. we guess that the domain wall may have the tunnel effect. In fact, not only the single cosmologic behaviour but also the anomaly, it all demonstrate the AdS space's specificity and the significance of this study. 

For the discussible works, when we know how the domain wall universe evolves in the charged AdS dilaton black hole background, the stability of universe is interested for us. Simplify, we don't think the fluctuation of bulk, while it just allows the domain wall to do perturbation \cite{Li:2018jxy}, \cite{Ong:2012yf}. Then we also want to study the cosmologic anomaly. If it is belong to a  quantum effect, will it have a relation with the Hawhing radiation? Also we want to calculate the observations in different cosmologic stage, such as, inflation, reheating and so all. This will provide  strong evidence for the existence of the domain wall universe and extra dimension.

\section*{Acknowledgements}
The work is supported by National Natural Science Foundation of China (No.11875081 and No.10875009).

\end{document}